# LABORATORY RESULTS AND STATUS UPDATE FOR PATHFINDER AT LBT, THE LINC-NIRVANA NGS GROUND-LAYER AO SUBSYSTEM

Derek Kopon[1,a], Al Conrad[1], Thomas Bertram[1], Tom Herbst[1], Martin Kürster[1], Jürgen Berwein[1], Roberto Ragazzoni[2], Jacopo Farinato[2], Valentina Viotto[2], Maria Bergomi[2], Ralf-Rainer Rohloff[1], Harald Baumeister[1], Fulvio De Bonis[1], Ralph Hofferbert[1], Alessandro Brunelli[2], Carmelo Arcidiacono[3], Jorg-Uwe Pott[1], Peter Bizenberger[1], Florian Briegel[1], Daniel Meschke[1], Lars Mohr[1], Xianyu Zhang[1], Frank Kittmann[1]

[1]Max-Planck-Institut für Astronomie, Königstuhl 17, D-69117 Heidelberg, Germany
[2]INAF, Osservatorio Astronomico di Padova, Vicolo Osservatorio 5, 35122 Padova, Italy
[3]INAF, Osservatorio Astronomico di Bologna, via Piero Gobetti 101, 40129 Bologna, Italy

**Abstract.** The full LINC-NIRVANA instrument will be one of the most complex ground-based astronomical systems ever built. It will consist of multiple subsystems, including two multi-conjugate ground layer AO systems (MCAO) [1] that drive the LBT adaptive secondaries [2, 3], two mid-high layer AO systems with their own Xynetics 349 actuator DM's , a fringe tracker, a beam combiner, and the NIR science camera. In order to mitigate risk, we take a modular approach to instrument testing and commissioning by decoupling these subsystems individually. The first subsystem tested on-sky will be one of the ground-layer AO systems, part of a test-bed known as the Pathfinder. The Pathfinder consists of a 12-star pyramid wavefront sensor (PWFS) that drives one of the LBT's adaptive secondaries, a support structure known as "The Foot," and the infrared test camera (IRTC), which is used for acquisition and alignment. The 12 natural guide stars are acquired by moveable arms called "star enlargers," each of which contains its own optical path. The Pathfinder was shipped from MPIA in Heidelberg, Germany to the LBT mountain lab on Mt. Graham, Arizona in February 2013. The system was unpacked, assembled in the LBT clean room, and internally optically aligned. We present the results of our system tests, including star enlarger alignment and system alignment. We also present our immediate plans for on-sky closed loop tests on the LBT scheduled for late Fall. Because plans for all ELTs call for ground layer correction, the Pathfinder provides valuable preliminary information not only for the full LINC-NIRVANA system, but also for future advanced MCAO systems.

## 1. Introduction

LINC-NIRVANA is the near-IR Fizeau interferometric beam combiner of the LBT that will achieve 10 mas resolution (in J-band) over a 10 arcsec field of view using two-layer multi-conjugate adaptive optics (MCAO) and fringe tracking [4]. The ground-layer wavefront sensor (GWS) for the DX side of the LBT, called the Pathfinder [5], is currently being commissioned at the telescope with first light scheduled for November 2013. Fig. 1 shows the Pathfinder mounted to the platform of the LBT.

The main components of Pathfinder are the GWS with 12 PWFSs [4, 5] that move independently to acquire natural guide stars, a blue steel support structure called "The Foot" that supports the GWS at the same height as the full LINC-NIRVANA optical bench, the IRTC, the electronics cabinet, and an annular mirror design to send a 2-4 arcmin annular field to the GWS (Fig. 3).

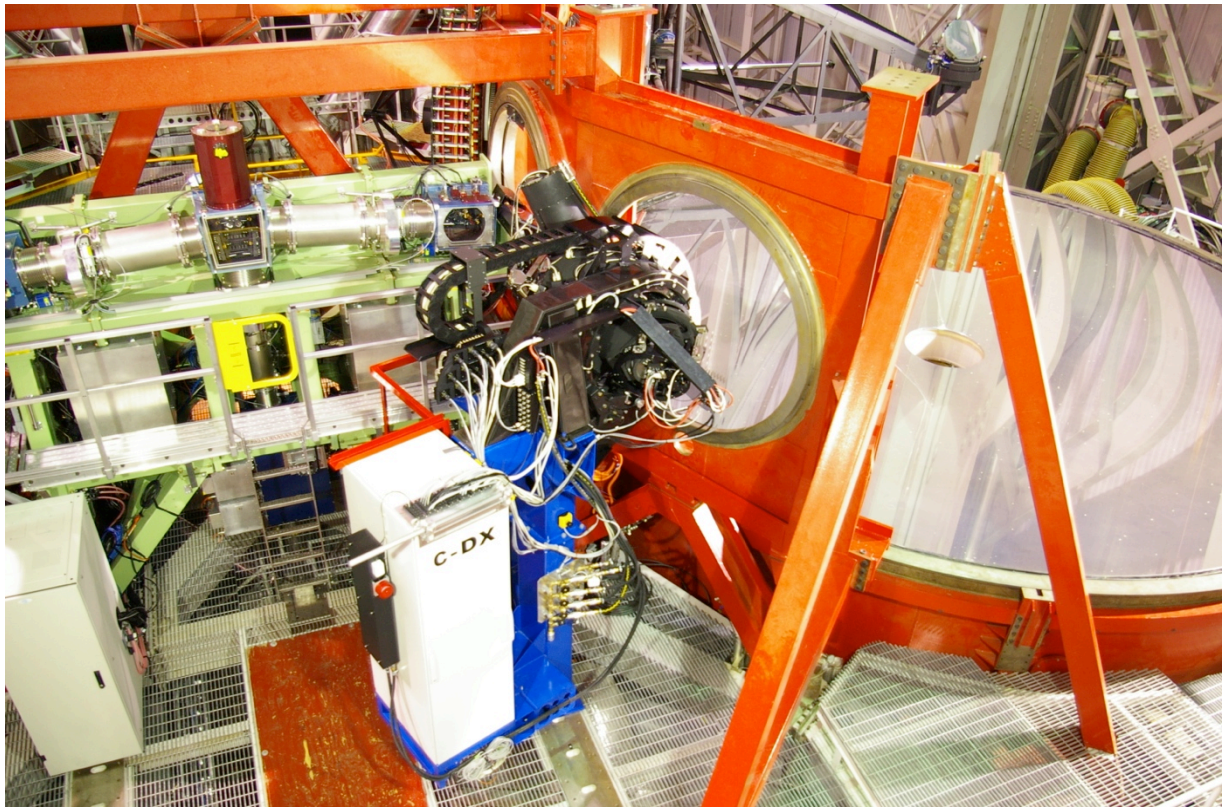

Fig. 1  Pathfinder at the LBT.

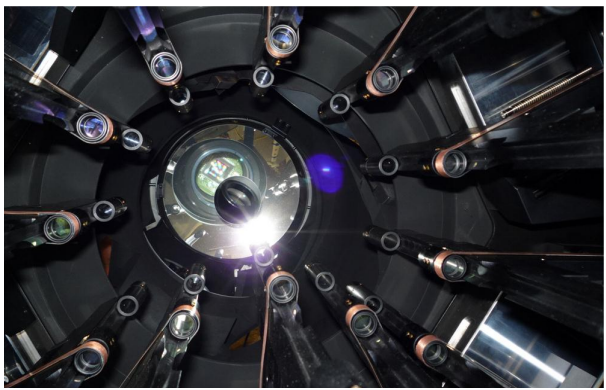
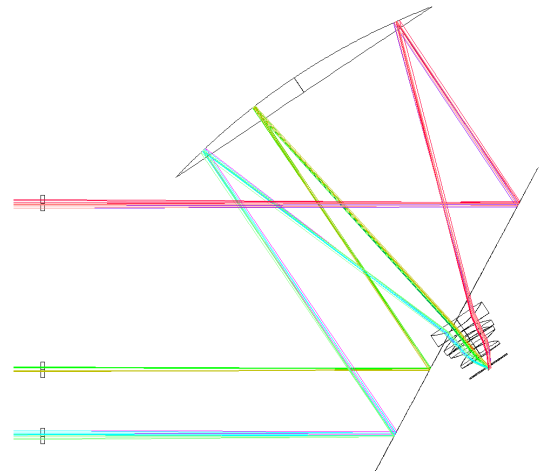

Fig. 2  The GWS with 12 moveable PWFSs to acquire natural guide stars at varying asterisms.  Left:  Photo of the GWS in the lab at MPIA.  Right:  Raytrace showing the light from three guide stars reflected off of a flat to a parabolic mirror before being optically coadded at the CCD50.

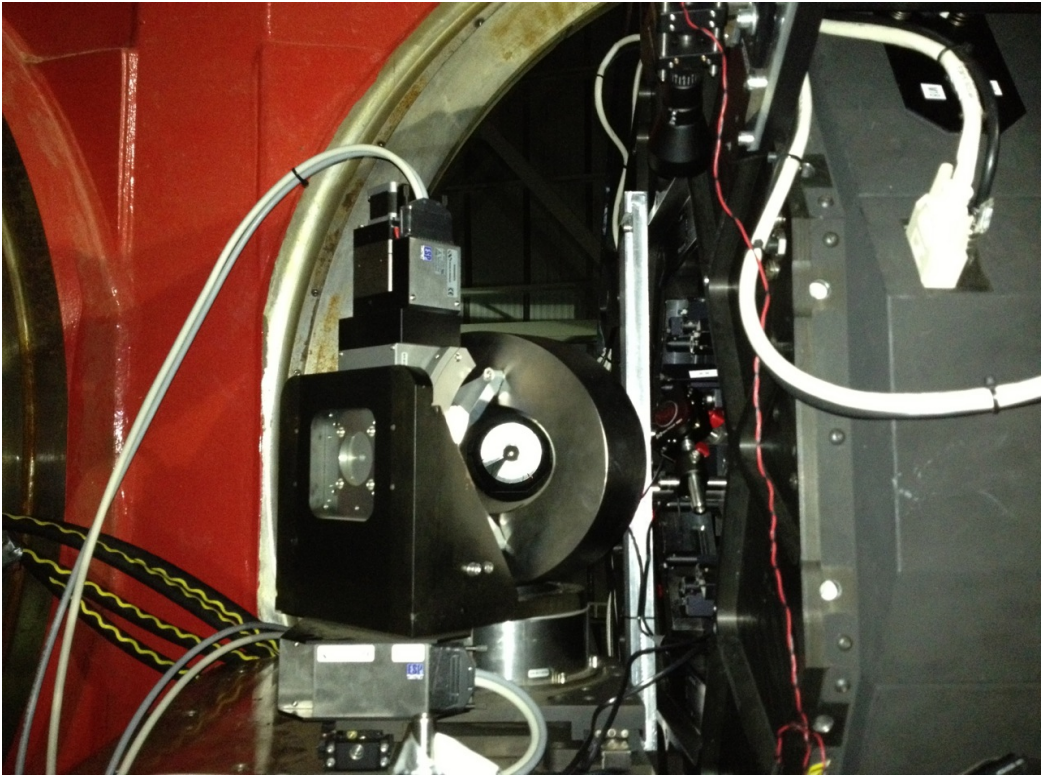

Fig. 3  Photograph of the annular mirror and the GWS.  The secondary mirror can be seen through the hole in the annular mirror.

## 2.   Current status

The 2013 commissioning plan for Pathfinder is shown in Table 1.  At the time of this writing, T1-T4 have been completed and T5 will begin in a few weeks.

Table 1 2013 Pathfinder Commissioning Plan

| | Tasks/Goals | Dates |
|---|---|---|
| T1 | Unpacking, assembly, and verification of rotator bearing accuracy and star enlarger accuracy.  Magic Lantern and annular mirror setup | March 1-15 |
| T2 | Star enlarger mapping and BCU software testing | April 3-11 |
| T3 | Craning of PF from the LBT mountain lab to our focal station on the telescope | April 24-28 |
| T3.5 | IRTC testing, fixturing, fit checking and setup of alignment tools | June 17-24 |
| T4 | Alignment of Pathfinder mechanical rotation axis to the telescope optical axis | Oct. 1-9 |
| T5 (day) | Interaction matrix calibration and test with off-sky closed loop test | Nov. 9-14 |
| T5 (night) | Acquisition, record pupil of star on CCD, send slopes | Nov. 15-16 |
| T6 (day) | Interaction matrix calibration and test with off-sky closed loop test | Dec. 3-8 |
| T6 (night) | Acquisition, record pupil of star on CCD, send slopes | Dec. 9-10 |

### 2.1. Shipping from Heidelberg and initial system assembly and test

In February 2012 the Pathfinder was shipped from the MPIA laboratory in Heidelberg, Germany to the LBT mountain lab.  The system was unpacked and assembled during the T1 commissioning run in March.  The wobble of the rotator bearing was measured using an auto collimating interferometer (Fig. 4) and found to still be in spec after the long trip from Heidelberg.  The star enlargers were precision aligned to the rotation axis of the bearing so their pupil images all optically coadd correctly on the CCD50 [6, 7]. The cooling manifold for the Pathfinder was installed and test, as were all motors, the annular mirror, and the "Magic Lantern" (ML) calibration source.

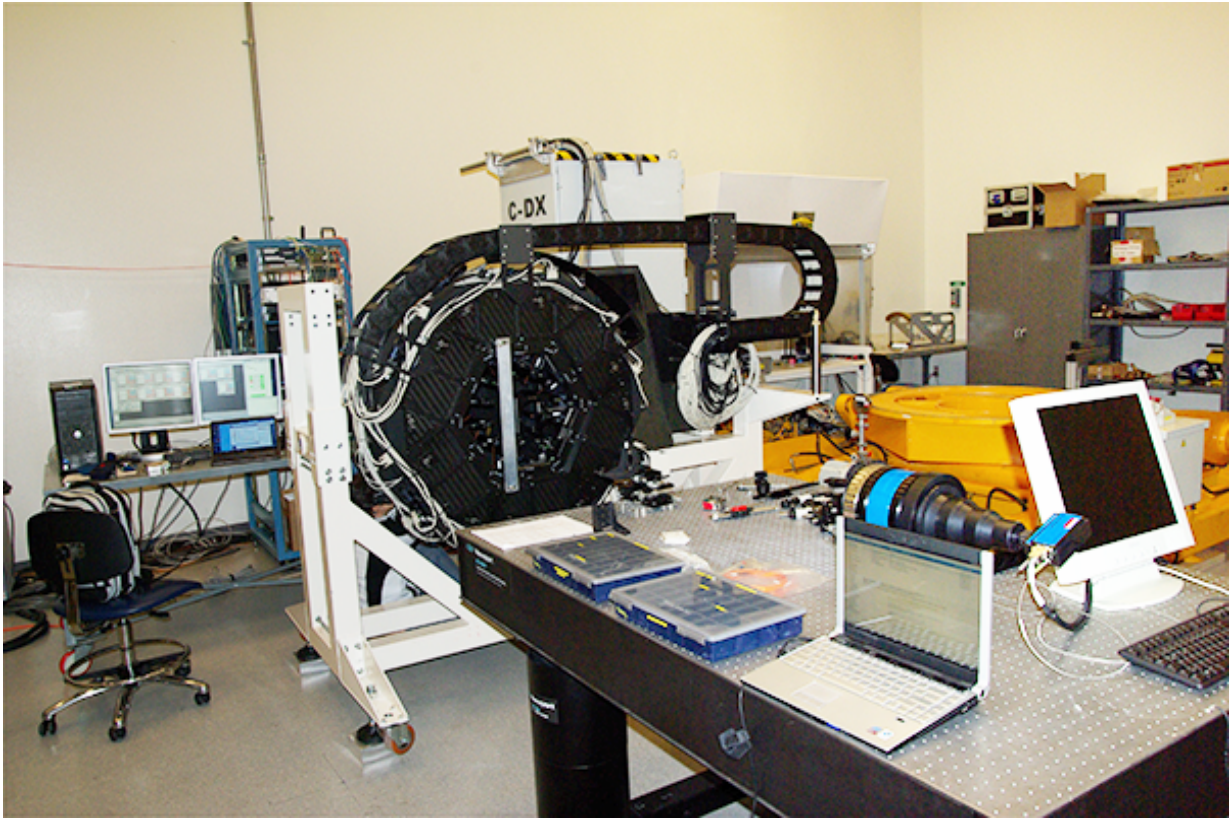

Fig. 4  The optical test setup for measuring rotator bearing "wobble." The GWS had not yet been craned to "The Foot," which occurred later in T1.

## 2.2. Star enlarger mapping and Magic Lantern tests

During the T2 commissioning run in April, the team mapped the coordinate system of the star enlargers using two different methods. The first was to photograph all of the star enlargers as each one was moved individually over its travel range using a stationary SLR camera. These photographs were then calibrated against a spatial reference of known size. The second method was to use the ML to simulate a point source on each star enlarger that produces pupil images on the CCD50 (Figure 5). The ML is an F/15 telescope simulator mounted on three precision translation stages that can move a simulated guide star by a known distance in the field. Known displacements of the simulated star were then used to map the motion of the star enlargers.

Also on T2, software interfaces with the LBT telescope control system were tested. Slopes were sent from the GWS to the LBT adaptive secondary. Other routines were also tested, such as a routine to automatically center the guide star on the tips of the star enlarger pyramids.

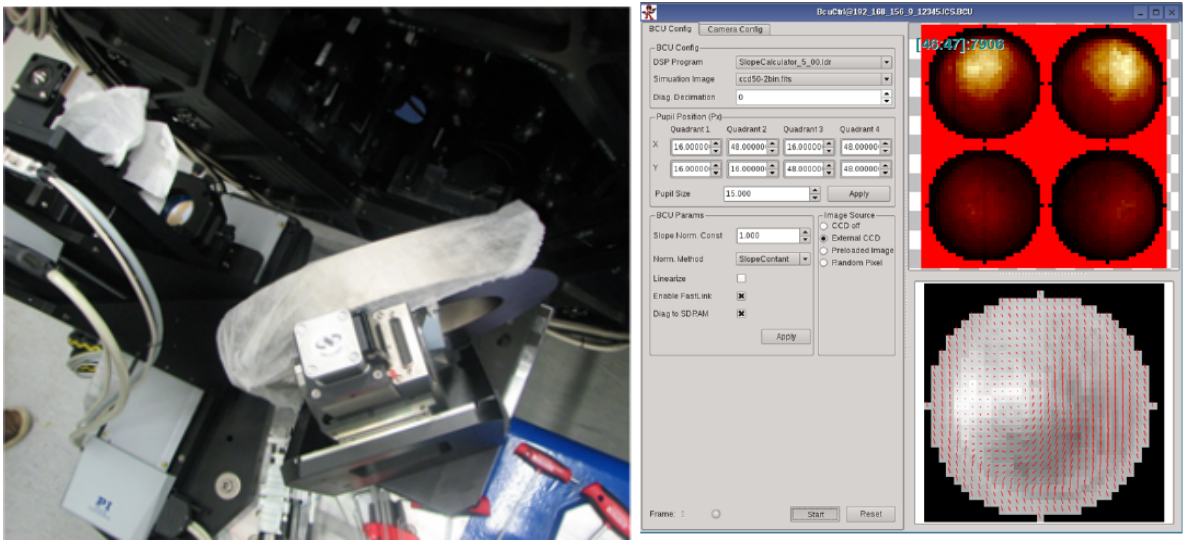

Fig. 5 Left: The "Magic Lantern" F/15 telescope simulator, the annular mirror, and the GWS. Right: Screen shot of the pupil images from the CCD50 created by the ML light source.

## 2.3. Pathfinder to telescope alignment

The alignment of the Pathfinder mechanical rotation axis to the optical axis of the DX side of the LBT presented a unique challenge because of the absence of any on-axis optical surfaces or fiducials due to the annular field of the instrument. To solve this problem, a rotating laser test was designed that involved mounting a laser on the GWS and reflecting it off the annular mirror to the tertiary and then to a translucent alignment screen (Fig. 8) positioned at the telescope prime focus. The Pathfinder bearing was then rotated, thereby tracing a circle with the laser on the alignment screen. The center of this circle defines a point in space through which the mechanical axis of the Pathfinder passes (Fig. 6).

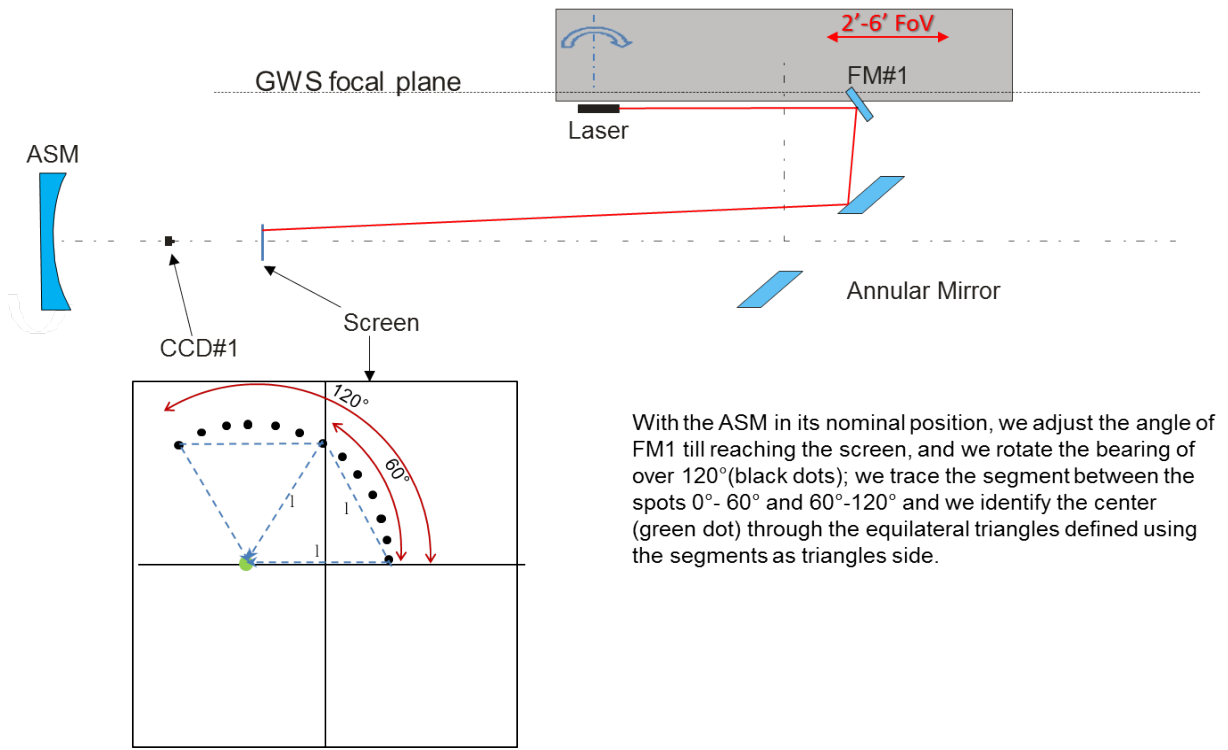

With the ASM in its nominal position, we adjust the angle of FM1 till reaching the screen, and we rotate the bearing of over 120°(black dots); we trace the segment between the spots 0°- 60° and 60°-120° and we identify the center (green dot) through the equilateral triangles defined using the segments as triangles side.

Fig. 6 Configuration 1 defines the mechanical axis of the Pathfinder rotation bearing.

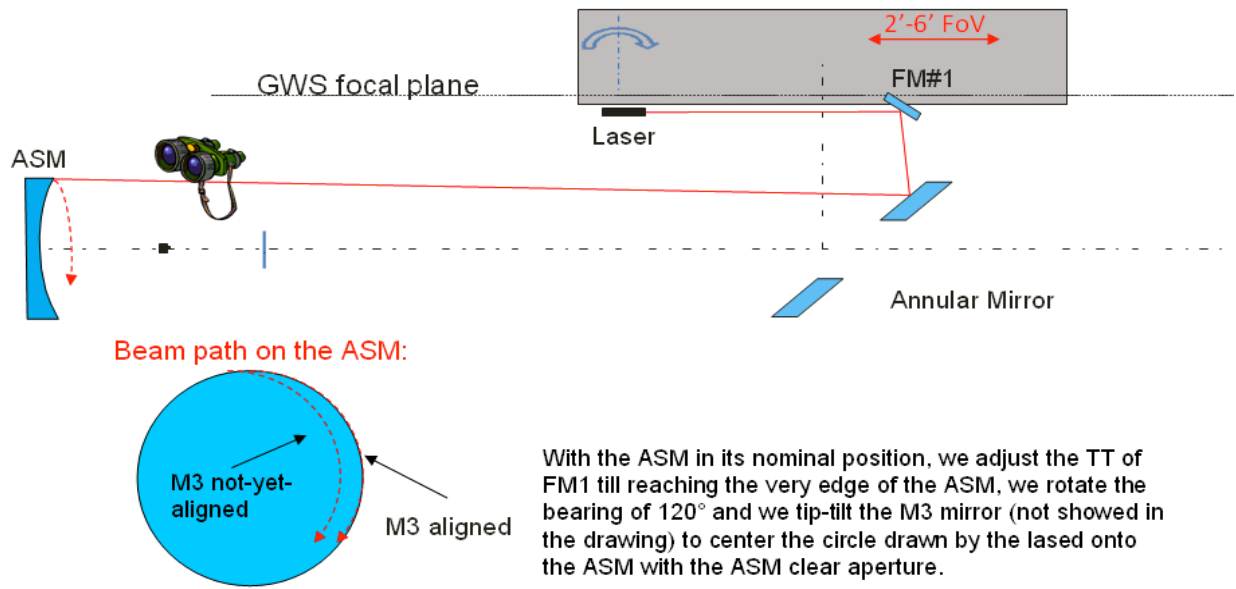

Fig. 7  Configuration 2 defines the optical axis of the LBT secondary mirror.

In a second configuration, the alignment laser is reflected from the GWS to the annular mirror to the tertiary to the secondary and then to the alignment screen <u>from the other side</u> (hence the need for a translucent screen).  The GWS bearing is again rotated tracing a circle with the laser on the alignment screen.  The center of this circle defines a point in space through which the optical axis of the telescope passes (Fig. 7).  By going back and forth between the first and second configuration of this test and iteratively adjusting the tip/tilt of the annular mirror and the tertiary, the centers of these two circles are made to be coincident and the mechanical axis of the Pathfinder is thereby aligned to the telescope optical axis.

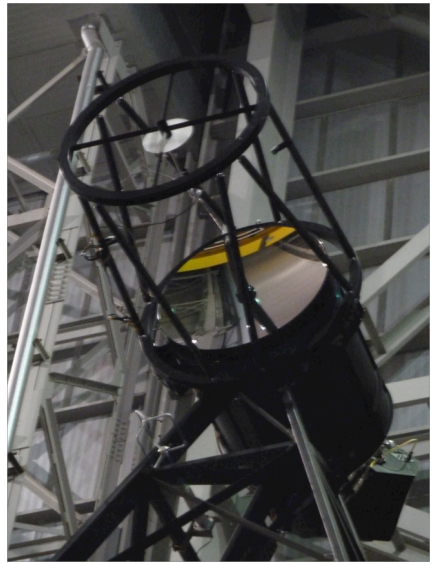

Fig. 8  Our translucent alignment screen mounted to the secondary, located at the prime focus.  A small video camera on a post is mounted a few inches from the alignment screen to give us a detailed view of the laser spots. This cage normally holds the LBT retroreflector that is used for calibrating the adaptive secondary mirror.

## 3.  Fall 2013 Commissioning Plan

The final two commissioning runs of 2013, T5 and T6, will both have as their objectives off-sky and on-sky closed loop tests.  Each run consists of approximately one week of daytime engineering time followed by two nights of technical time.

## 3.1. Off-sky closed loop test

We have designed an off-sky double pass closed loop test similar to that used by the LBT FLAO system that will be used to run the Pathfinder AO system closed loop off-sky in order to calibrate the interaction matrix. As our light source, we will use the Magic Lantern calibration source with a few modifications. The stop of the ML will be enlarged to produce a beam faster than the telescope F/15 beam so that we will over-fill the secondary. The ML will launch this ~F/11 beam through a beamsplitter to the tertiary and secondary and then into the retroreflector (RR), an on-axis paraboloid and flat pair (the same used by the other AO teams at LBT). The return beam from the RR will reflect off of the beamsplitter to another beamsplitter that will send part of the light to a technical viewing camera located in a focal plane and the rest to the pyramid tip of on of the star enlargers. The technical viewer will be used to image the spot and give a visual measure of gross coma that would saturate the PWFS and can be aligned out. Then the GWS will be used to close the loop on the whole test with the secondary mirror. As with the FLAO system, artificial turbulence can be injected into the secondary in order to simulate our on-sky performance.

## 3.2. On-sky closed loop test

With our off-sky calibrated interaction matrix we will close the loop on natural guide stars during our nights in both T5 and T6. In doing so, we will gain experience acquiring one natural guide star, followed by acquiring two or more simultaneously. As we acquire more guide stars, the Strehl over our field becomes better and more uniform (Fig. 9).

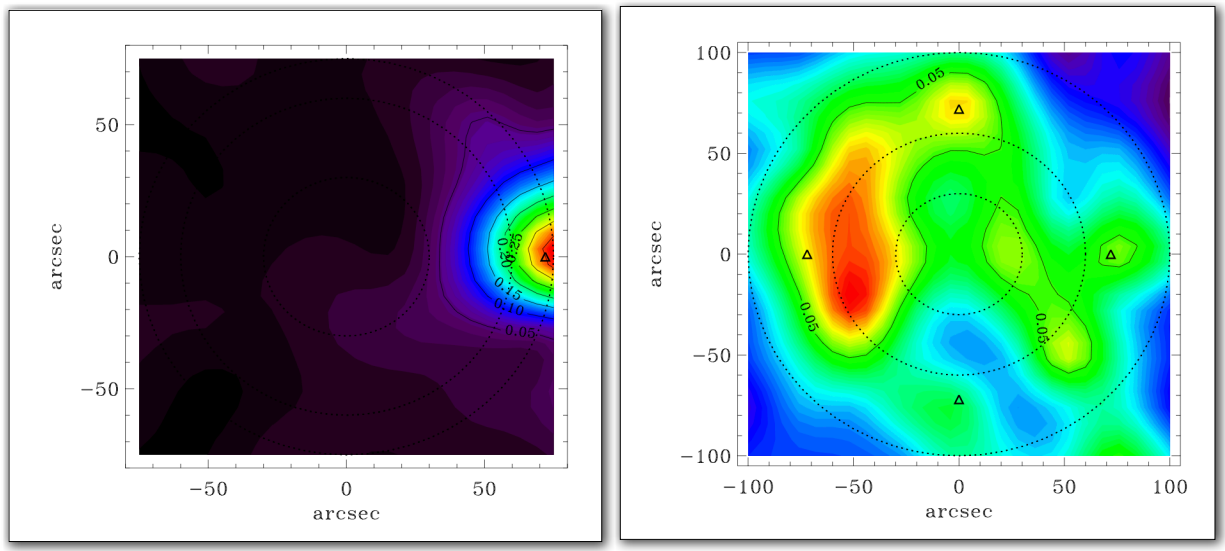

Fig. 9 Left: Simulated Strehl with one guide star over a 160 arcsec FOV. Right: Simulated Strehl for four guide stars over a larger (200 arcsec) FOV.

# 4. Conclusion

The Pathfinder, the first LINC=NIRVANA subsystem at the LBT, proceeds through the 2013 commissioning schedule towards on-sky closed loop ground-layer AO in the next few months.

# 5. References

1.   R. Ragazzoni, J. Farinato, and E. Marchetti, "Adaptive optics for 100-m-class telescopes: new challenges require new solutions," *Proc. SPIE*, vol. 4007. pp. 1076–1087, 2000.

2.    A. Riccardi, M. Xompero, R. Briguglio, F. Quirós-Pacheco, L. Busoni, L. Fini, A. Puglisi, S. Esposito, C. Arcidiacono, E. Pinna, P. Ranfagni, P. Salinari, G. Brusa, R. Demers, R. Biasi, and D. Gallieni, "The adaptive secondary mirror for the Large Binocular Telescope: optical acceptance test and preliminary on-sky commissioning results," *Proc. SPIE*, vol. 7736. p. 77362C–77362C–12, 2010.

3.    S. Esposito, A. Riccardi, L. Fini, A. T. Puglisi, E. Pinna, M. Xompero, R. Briguglio, F. Quirós-Pacheco, P. Stefanini, J. C. Guerra, L. Busoni, A. Tozzi, F. Pieralli, G. Agapito, G. Brusa-Zappellini, R. Demers, J. Brynnel, C. Arcidiacono, and P. Salinari, "First light AO (FLAO) system for LBT: final integration, acceptance test in Europe, and preliminary on-sky commissioning results," *Proc. SPIE*, vol. 7736. pp. 773609–773612, 2010.

4.    R. Hofferbert, H. Baumeister, T. Bertram, J. Berwein, P. Bizenberger, A. Böhm, M. Böhm, J. L. Borelli, M. Brangier, F. Briegel, A. Conrad, F. De Bonis, R. Follert, T. Herbst, A. Huber, F. Kittmann, M. Kürster, W. Laun, U. Mall, D. Meschke, L. Mohr, V. Naranjo, A. Pavlov, J.-U. Pott, H.-W. Rix, R.-R. Rohloff, E. Schinnerer, C. Storz, J. Trowitzsch, Z. Yan, X. Zhang, A. Eckart, M. Horrobin, S. Rost, C. Straubmeier, I. Wank, J. Zuther, U. Beckmann, C. Connot, M. Heininger, K.-H. Hofmann, T. Kröner, E. Nussbaum, D. Schertl, G. Weigelt, M. Bergomi, A. Brunelli, M. Dima, J. Farinato, D. Magrin, L. Marafatto, R. Ragazzoni, V. Viotto, C. Arcidiacono, G. Bregoli, P. Ciliegi, G. Cosentino, E. Diolaiti, I. Foppiani, M. Lombini, L. Schreiber, F. D'Alessio, G. Li Causi, D. Lorenzetti, F. Vitali, M. Bertero, P. Boccacci, and A. La Camera, "LINC-NIRVANA for the large binocular telescope: setting up the world's largest near infrared binoculars for astronomy," *Opt. Eng.*, vol. 52, no. 8, p. 81602, 2013.

5.    A. R. Conrad, C. Arcidiacono, H. Baumeister, M. Bergomi, T. Bertram, J. Berwein, C. Biddick, P. Bizenberger, M. Brangier, F. Briegel, A. Brunelli, J. Brynnel, L. Busoni, N. Cushing, F. De Bonis, M. De La Pena, S. Esposito, J. Farinato, L. Fini, R. F. Green, T. Herbst, R. Hofferbert, F. Kittmann, M. Kuerster, W. Laun, D. Meschke, L. Mohr, A. Pavlov, J.-U. Pott, A. Puglisi, R. Ragazzoni, A. Rakich, R.-R. Rohloff, J. Trowitzsch, V. Viotto, and X. Zhang, "LINC-NIRVANA Pathfinder: testing the next generation of wave front sensors at LBT," *Proc. SPIE*, vol. 8447. p. 84470V–84470V–10, 2012.

6.    R. Ragazzoni, "Pupil plane wavefront sensing with an oscillating prism," *J. Mod. Opt.*, vol. 43, pp. 289–293, 1996.

7.    R. Ragazzoni, A. Ghedina, A. Baruffolo, E. Marchetti, J. Farinato, T. Niero, G. Crimi, and M. Ghigo, "Testing the pyramid wavefront sensor on the sky," *Proc. SPIE*, vol. 4007. pp. 423–430, 2000.

8.    R. Ragazzoni, E. Diolaiti, J. Farinato, E. Fedrigo, E. Marchetti, M. Tordi, and D. Kirkman, "Multiple field of view layer-oriented adaptive optics," *A&A*, vol. 396, no. 2, pp. 731–744, 2002.

9.    E. Marchetti, N. N. Hubin, E. Fedrigo, J. Brynnel, B. Delabre, R. Donaldson, F. Franza, R. Conan, M. Le Louarn, C. Cavadore, A. Balestra, D. Baade, J.-L. Lizon, R. Gilmozzi, G. J. Monnet, R. Ragazzoni, C. Arcidiacono, A. Baruffolo, E. Diolaiti, J. Farinato, E. Vernet-Viard, D. J. Butler, S. Hippler, and A. Amorin, "MAD the ESO multi-conjugate adaptive optics demonstrator," *Proc. SPIE*, vol. 4839. pp. 317–328, 2003.